# AN OVERVIEW OF DEEP LEARNING IN MEDICAL IMAGING

Imran Ul Haq

College of Information Sciences and Technology, Northwest University, Xi'an 710069, China

**Abstract:**

Machine learning (ML) has seen enormous consideration during the most recent decade. This success started in 2012 when an ML model accomplished a remarkable triumph in the ImageNet Classification, the world's most famous competition for computer vision. This model was a kind of convolutional neural system (CNN) called deep learning (DL). Since then, researchers have started to participate efficiently in DL's fastest developing area of research. These days, DL systems are cutting-edge ML systems spanning a broad range of disciplines, from human language processing to video analysis, and commonly used in the scholarly world and enterprise sector. Recent advances can bring tremendous improvement to the medical field. Improved and innovative methods for data processing, image analysis and can significantly improve the diagnostic technologies and medicinal services gradually. A quick review of current developments with relevant problems in the field of DL used for medical imaging has been provided. The primary purposes of the review are four: (i) provide a brief prolog to DL by discussing different DL models, (ii) review of the DL usage for medical image analysis (classification, detection, segmentation, and registration), (iii) review seven main application fields of DL in medical imaging, (iv) give an initial stage to those keen on adding to the research area about DL in clinical imaging by providing links of some useful informative assets, such as freely available DL codes, public datasets Table 7, and medical imaging competition sources Table 8 and end our survey by outlining distinct continuous difficulties, lessons learned and future of DL in the field of medical science.

DL influences any area of public and private life and is becoming an essential part of daily life. The strength of DL comes from its ability to mimic the actions of human brain neurons in the neocortex, where the mechanism of thinking occurs. DL also attempts like brain neurons to understand and identify patterns in digital images. The strength of DL depends on the number of neuron layers used for computation. The process strengthened using Graphics Processing Units (GPUs) with high-speed processor.

**Keywords:** CNN, Deep learning, Medical Image analysis, Segmentation, Classification, Detection, Review

## 1. INTRODUCTION

As long as medical images can be processed and loaded into a computer, researchers have developed automatic processing systems. At the beginning (1970-1990), medical image analysis was conducted with the sequential application of mathematical modeling and simple pixel processing to create composite rule-based models that solved unique tasks. During the late 1990s, supervised approaches of ML were becoming ever more common for analysis of medical images, where a model is built and trained using a training set of data. This pattern recognition method is still famous, making the foundation of several effective medical imaging systems. In this way, we have seen a change from entirely human-designed systems to computer-trained systems using example data extracting of feature vectors. Human re-searchers are still extracting distinguishing features from the images with handcrafted models. The next sensible move was to empower computers to know data representing features in an optimized way for the problem we want to solve. Many DL algorithms are based on this concept. DL models (networks) are made up of several layers that convert input data into outputs for learning high-level features. To date, CNNs are the most popular class of image analysis models, and Deep convolutional networks are currently the strategy of choice in computer vision. The application of DL in health care is still in its development. However, there are many robust research programs, and several major corporations are undertaking ML-based healthcare projects.

ML algorithms can be actively involved in all areas of medicine, and thus greatly improving the way of treatment and practiced. Recently, the popularity of ML systems in the computer vision field came at a period of great success when medical records are being rapidly digitalized. Medical images form a vital part of the electronic health records (EHR) of a patient. Human radiologists examine these records through the traditional

Corresponding author: Imran Ul Haq (imhaq123@stumail.nwu.edu.cn)

way. As humans, all of them are constrained by speed, exhaustion, and expertise. Training a professional radiologist takes years at a considerable financial cost. Overdue or incorrect treatment would harm the patient. For this reason, an automatic, precise and efficient ML algorithm is suitable for medical image processing.

There are countless modalities of imaging, and the extent of their use are growing. In the period between 1996 and 2010, a study investigated the use of imaging across 6 large interconnected health-care medical systems in the United States. They wrote that 30.9 million medical imaging tests were taken in that period [1]. The researchers observed that the use of magnetic resonance imaging (MRI), computed tomography (CT) and positron emission tomography (PET) rose 10%, 7.8%, and 57% respectively, during the study period.

Digital imaging modalities used for medical diagnostics include ultrasound (US), X-ray, MRI and CT scans, PET scans, histology slides, retinal photography and dermoscopic images. Some of the above methods investigate different organs (like MRI & CT), and others are limited to organs (dermoscopy & retinal photography). There is also a difference in the amount of data produced from each report. The size of a slide for histology may be an image file of some megabytes., but an MRI file can be several hundred megabytes. It has technical consequences for how we pre-process the data, and for the architecture design of an algorithm considering processor and memory constraints. In this study, we concentrate on DL models in particular that were applied to medical images.

### 1.1 Deep learning

The traditional models of ML, we use manually crafted features to train our network to perform the desired task. These manually crafted features are derived from some raw data or acquired from other basic ML methods. In the case of DL, the features are automatically derived and learned by computers from the data, avoiding the challenging phase of the manual derivation of features. These days the most popular models of DL are the numerous types of Artificial Neural Networks (ANN), but some other types are also available. The key feature that distinguishes DL approaches from other ANN methods is that they emphasize feature learning, which in simple words means that they automatically learn features from data. The discovery of features and the execution of tasks are combined, and thus both improve during the single phase of training. The interest in DL in the field of medical imaging caused by CNNs, an efficient means of learning valuable image representations and other structured data. As long as CNNs not be used effectively, these features were designed manually or generated with the help of less efficient ML techniques. In CNNs, some powerful standards are encoded that allow us to understand how they are efficient. An overview of DL models and the building blocks of CNNs are discussed in the following sections.

## 2. OVERVIEW OF DL MODELS

In this section, we will discuss the DL models used in the field of medical image analysis covered by the reviewed articles. ML uses statistical tools to classify data into two or more groups by learning from the data. Two types of DL models are supervised and unsupervised. In the first case, the input data and the labels corresponding to input data are given to the system. The system links the given input data with the corresponding labels. In the case of unsupervised models, input data without tags are provided to the system. The system looks and searches for some patterns in the training data. These maps or patterns are used to give output for the unseen data. The DL models discussed below have been divided into two categories of Supervised and Unsupervised DL models.

### 2.1 Supervised Learning Models

### 2.1.1 Convolutional Neural Networks

CNNs are the most studied algorithms for ML in medical image analysis.[2] The cause behind that when the input images altered, CNNs maintain spatial relationships. Spatial relations are vital in radiology, for instance, how the bone edge joins the muscle or where normal lung tissue interacts with cancerous tissue. A CNN takes a raw pixel input image and transforms it through convolutional layers, RELU (Rectified Linear Unit) linear layers and pooling layers. The output is given to a final fully connected layer that assigns a final class to the image for classification problem as shown in Figure 1.

### a) Convolution Layer

The name of this layer is taken from the operation applied on this layer called the convolution. On this layer we convolve input and kernel. Input tensor in image processing contains the input or pixel values, which are taken from a particular location of the image. Kernel is a small array of numbers applied to the input tensor. The corresponding output called feature map, is obtained by applying dot operation between the kernel and input tensor. Then, the filter is moved to the next location on the image according to the stride length and repeat these steps again to cover the whole image. The process is repeated with multiple kernels to obtain different features maps, that represent different characteristics of the input tensor.[3] Two important hyperparameters for the convolution operation are the number and size of the kernel. The most commonly used sizes are 3 x 3, 5 x 5 and 7 x 7, while the number of kernels is arbitrary and shows the depth of the feature map. When we apply the convolution operation as described above, the center of each kernel does not overlap the outermost element of the input tensor which causes decrease in the height and width of the output feature map compare to input tensor, in every succeeding layer. Padding can be used to overcome this problem, where rows and columns of zeros are added to the input tensor, that enable kernels to overlap the outermost elements of the input tensor, resulting the same dimension throughout the convolution process. Another important hyperparameter for convolution is the stride, which is the distance between the two successive kennel's positions. Most common choice for stride is 1, but sometimes values greater than 1 are also used for the purpose of down sampling. Convolution takes advantage of the following three intrinsic ideas for performing computationally competent in ML: sparse connections, parameter or weight sharing and representation equivariant.[4] Unlike other NNs where each output neuron is connected to every input neuron of the next layer, CNN neurons are sparsely connected, which means that just some of the input neurons are linked to the subsequent layer. Due to the small receptive field, it is possible to learn meaningful features and the respective numbers of weights for measurement can be reduced rapidly, which increases the efficiency of the algorithm. CNNs reduce the memory storage requirements by using every filter with its set weights in various places across the entire image called parameter sharing. The weights consequences are diverse across layers in a fully connected NN, which are discarded after using a single time. Parameter sharing increases the value of equivariant representation. Which implies, the input translations lead to a map translation of the corresponding feature.

The "*" symbol represents the operation of convolution. The convolution between the input I(t) and kernel K(a), is represented by the following expression, S(t) is the output:

$$S(t) = (I * K)(t)$$

The discretized convolution in one dimension, where t can take only integer values is given by:

$$S(t) = \sum_a K(a) \cdot I(t - a)$$

In case we have a two-dimensional input image I (M, N) and kernel K (m, n), where M, m and N, n are the number of rows and columns of image and kernel, then the size of their convolution S(i, j) will be (M - m + 1) x (N - n + 1) and can be expressed as below:

$$S(i, j) = \sum_{k=1}^{m} \sum_{l=1}^{n} I(i + k - 1, j + l - 1) K(k, l)$$

where i and j take values from 1 to M - m + 1 and 1 to N - n + 1 respectively. As described above this implementation produces output of decreased size at every successive layer, which can be solve by padding.

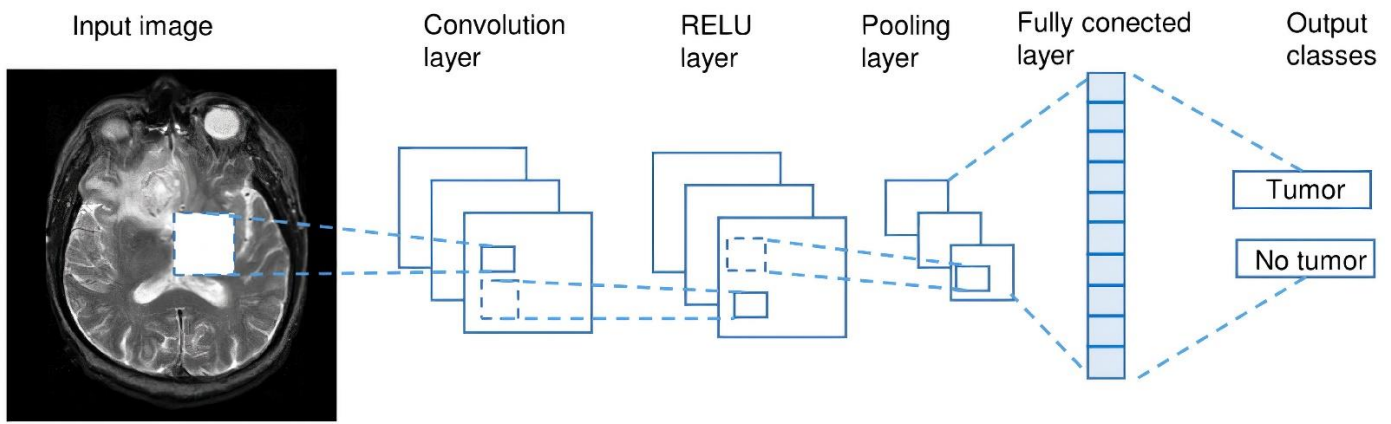

Fig 1: An input image of a diseased axial slice of a T2-weighted brain MRI is passed through a graphical representation of a CNN. The input image feature extraction is conducted through the Convolution, RELU and pooling layers before classification by the fully connected layer [5]

### b) *Rectified Linear Unit (RELU)*

The RELU layer is a function of activation, which makes every negative input equal to zero. Because of which the training process and other calculations speed up and helps elude the gradient issue. If x is the neuronal input, this function can be represented by the following formula:

$$f(x) = max\ (\ 0, x\ )$$

Other activation functions include sigmoid, tanh, leaky RELUs, Randomized RELUs, and parametric RELUs.

### c) *Pooling Layer (PL)*

The purpose of this layer is to minimize the number of parameters that need to be measured and reduce the input image size by reducing its breadth and height without changing its depth. Max-pooling is the most widely used. It only picks the filter's most considerable input value, and all other values are discarded, resulting in the most potent activation over a neighborhood. L2-normalization and average pooling include different types of PL.

### d) *Dropout:*

A fundamental concept that makes a significant improvement to CNNs achievements, is that it combines several models in a composite one to achieve higher efficiency than using an individual model. Dropout is a common strategy based on NN's stochastic sampling. By randomly deleting neurons during training, every batch of training data end using different networks, and the weights of the network qualified are adjusted based on optimizing several network variants.

### e) *Fully Connected Layer (FCL)*

The last component of CNN architecture is FCL. The name implies that every single neuron of FCL is linked with every single neuron of the preceding layer. Just like the other layers, one or more FCLs can be present in the NN depending on the level of features wanted. FCL takes the input data from the previous layer and calculates the probability score for the various categories presented for classification. Essentially, FCL monitors the most strongly enabled features and their combination, which suggests that the picture belongs to a specific category. For example, cancerous cells hold a high ratio between DNA and cytoplasm than normal cells on histology glass slides. If DNA characteristics were strongly detected by the prior layer, CNN would be more able to predict the existence of cancerous cells. Together with backpropagation and stochastic gradient descent, methods used for NN training can help the CNN to select important features from training data.[6]

### 2.1.2 Transfer Learning

In contrast to general image identification, there is a lack of broadly labeled training datasets for medical image analysis. This problem was seen by comparing two datasets of Kaggle 2017 Data Sciences [7] for tumor detection in CT scans consist of 2000 patient's data and the ILSVRC 2017 dataset of general images which has more than a million images of thousand different object categories.[8] In the transfer learning approach, a moderately related or unrelated dataset along with a labeled dataset is used for training of the ML approach to evade the problem of insufficient amount of training data. The weights learned by CNN during the training on one dataset are then transferred to another CNN. This CNN is further trained by labeled medical data using the learned weights. Excluding the last FC layer, these weights may be applied to all or part of the CNN layers. While the use of transfer learning techniques along with CNNs is common in medical imaging. That can be extended to other general ML algorithms.

Shin et al.[9] investigated the effect of transfer learning and CNN for identifying grown lymph nodes of the abdomen and for the classification of interstitial lung disease using CT scans. Although the natural images were different from the medical images transfer learning was found beneficial. Ravishankar et al.[10] investigated the task of automatically converging a kidney's presence on US images. They used pre-trained CNN on ImageNet and concluded that, as we increase the degree of transfer learning, the CNN performance will improve. Tajbakhsh et al.[11] tested the efficacy of transfer learning across 3 imaging modalities in 4 applications: identification of pulmonary embolus on CT pulmonary angiograms, segmentation of the carotid artery wall layers on ultrasonic scans, identification of polyps on colonoscopy images and classification of colonoscopy video frames. They noticed that transferring more layers of learning increased CNN output than training a CNN from scratch. Unlike other computer vision tasks, for which shallow fine-tuning of the last few layers are enough, deep tuning of more layers is required in medical image analysis. They also noticed that different applications require different numbers of optimally trained layers. A DL framework was proposed by [12] that used a combination of TL and RNN for detection of children pneumonia. To solve the problem of model overfitting due to insufficient data, the model parameters trained on large datasets were used to initialize the model using TL. They got a recall rate of 96.7% and f1-score of 92.7% on classification of children pneumonia.

Nishio et al. [13] designed a CAD system for ternary classification of the primary lung cancer, metastatic lung cancer and benign nodule, and measured the effectiveness of the DCNN for CAD and the efficiency of TL. CAD system was tested with and without TL for the DCNN system using the VGG-16 CNN. The highest average accuracy they got for the validation set was 62.4% for the DCNN system without TL and 68.0% for the DCNN system with TL. For the proposed CAD system DCNN performed better than the conventional method and the DCNN's accuracy was further improved by using TL. Compare to ImageNet architecture, a lightweight and relatively simple model performed comparably when evaluated the performance for two medical imaging tasks, by using TL technique in .[14] Unlike other computer vision tasks, for which shallow fine-tuning of the last few layers are enough, deep tuning of more layers is required in medical image analysis. They also noticed that different applications require different numbers of optimally trained layers.

[15] used two deep TL models based on deep CNN which were trained on ImageNet dataset for binary and multiclass classification of breast tumors and improved the accuracy. The models were tested for magnification dependent and independent classification modes with optimized parameters, and reported an accuracy up to 98\% for multiclass and up to 100\% accuracy for binary classification. A systematic literature review was presented by [16] by comparing the healthcare experts and various TL techniques in detecting various diseases using medical images. They considered research articles from 2014 to 2019 based on their criteria, and concluded that recent advances in TL have achieved comparable performance with medical professionals in various fields.

### 2.1.3 Recurrent Neural Networks (RNNs)

Because of the text production ability of RNNs, their use is more common for sequential data analysis [17]. RNNs were used in different text processing projects, like speech recognition, machine translation, text prediction, language processing and image caption generation [18]. In simple RNN, a layer output is applied for next input, which is again provided to that layer, causing spatial memory capability. Simple RNNs have been changed into Gated Recurrent Units (GRUs) and Long Short-Term Memory (LSTM) networks to avoid waning gradient issues caused by backpropagation over time. These RNN modifications retain long-term dependencies to forget or delete some evidence accumulated. RNNs were used primarily for segmentation in the medical image investigation. Chen et al.[19] used a combination of RNN and CNN for segmentation of fungal and neuronal structures from 3D electron microscope images. The segmentation of 3D electron microscopic images of neurons and brain MRIs used as a multidimensional LSTM [20]. Shin et al.[21] identified X-ray images that are annotated with captions learned on radiology studies.

## 2.2 UNSUPERVISED LEARNING MODELS

In this section we will discuss unsupervised models like Auto encoder (AE), Generative Adversarial Networks (GANs) and Restricted Boltzmann Machine (RBM).

### 2.2.1 Autoencoders

Autoencoders (AEs) use an unsupervised strategy without labeled the input data, also called codings, to learn feature representation. The input data are given to the model; the model extracts codes from it and then reconstructs output data using these codes. The output data are also called reconstruction. The simple logic of AEs is that the input and output data should be as alike as possible; that is, AEs have a cost function that punishes the model in the case output different from the input. There are several useful features to AEs. First, they are used as a feature detector that does need training labels for the training data and can learn coding in an unsupervised way. Next, they minimize the dimensionality and difficulty of the model because coding mostly happens in a lower dimension. Thirdly, AEs create new output that is identical to the input data used for training. Such features are beneficial when processing medical images, where properly labeled data for training purposes is limited. One unique architectural feature is that the input and output layers of the AE have the same quantity of neurons. AEs, just like CNNs, also consist of hidden layers accumulated. The architecture of Stacked autoencoders (SAEs) is generally in a symmetrical shape, in which the reflective line passes from the hidden layer. Few strategies for optimizing AE output involve binding the decoder layer's weights to the encoder layer, training various AE sub-sets separately until stacking all and transfer learning [22]. The model accuracy can't be increased by merely stacking an increasing number of layers as the model can wind up doing a meaningless job of only photocopying input in the output. Which in simple words is, during training, the model achieves good results, but no useful feature representations have been learned that allows for the generalization and implementation of the model outside the training data.

Forcing models to learn useful representations involves the introduction of constraints. An example explains by Vincent et al.[22], called Denoising Autoencoder, in which the early hidden layers add Gaussian noise. The same purpose can be achieved by randomly turning off some neurons in the beginning hidden layers. Sparse AEs are the $2^{nd}$ example used to deactivate a certain fraction of the hidden layer's neurons and put them to nil value. When the number of active neurons were above a given threshold value, the selected cost function will penalize the model for achieving the goal. As Bengio states, for a given observation only a limited number of factors are essential, so the rest of the extracted features might be put equal to zero [23]. Kallenberg et al. merged supervised layers with unsupervised convolution layers trained as AE for mammogram classification into various thicknesses and shapes [24]. Classification of shapes was implied to assign whether a mammogram was depicted by breast cancer or not. They used a dataset composed of 2700 mammograms collected from three datasets. Before the input was given to a SoftMax classifier, a sparse autoencoder was used for parameter learning of the feature extracting convolution layers. The AUC score got by this Convolution stacked autoencoder (CSAE) model for the cancer classification task was 0.57, which was reported as the best by the authors.

Kingma and Welling defined an evolving architecture based on unsupervised learning called Variational Autoencoders (VAEs) [25]. VAEs are generative models consisting of decoder and encoder networks (Bayesian inference encoder). Stochastic gradient descent can be used for training the VAEs. Input data are mapped to latent space by the encoder network, approximating their right distribution via a Gaussian distribution. Latent space is a map back to output data by decoder network, guided and trained by two cost functions: the Kullback_Leibler divergence and a reconstruction loss function.

### 2.2.2 Generative Adversarial Networks (GAN)

GANs [26] are unsupervised learning models used for tasks related to medical image analysis. A GAN is a generative design, and in this sense, is close to a VAE. GANs consist of two competing models trained simultaneously and, like CNNs multilayer perceptron. These two models are just like two players rivaling in a game of zero-sum. Among these two CNNs, one generates images from artificial training and is named a generator. In comparison, the second CNN categorizes images from the generator if they are training images or artificial ones, is called a discriminator. The optimal outcome of this adversarial system is when the discriminator is unable to discriminate the real image from a generated one. In this situation, the likelihood of adding an item to any given set is equal i.e., 0.5. one plus point is that it is possible to train both CNNs (generator, discriminator) using techniques of dropout and backpropagation, without hefty deduction and Markov chains. Compare to other DL techniques; GANs are new, still have specific uses in the generation of synthetic medical data and brain MRI segmentation [27-29].For more details on the use of GANs in medical imaging please read [30].

### 2.2.3 Restricted Boltzmann Machines (RBMs)

Ackley et al.[31] invented Boltzmann machines in 1985, and a year later, Smolensky [32] updated it as Restricted Boltzmann Machines (RBMs). They are probabilistic, bidirectional graphical models. Just like other deep networks, RBMS also have visible and hidden layers which can be given as $\mathbf{v} = (v_1, v_2, \ldots, v_N)$ and $\mathbf{h} = (h_1, h_2, \ldots, h_M)$ respectively [4]. These layers are linked to each other, but the layers themselves do not have any connection. RBMs use the input data backward pass to produce a reconstruction and approximate the input data probability distribution. The bidirectional links connect the nodes in RBMs, such that one may get the latent interpretation of the feature h provided an input vector v and vice versa. In similarity with physical processes, the function of energy E related to a specific condition (v, h) of input and hidden elements is given by:

$$E(\mathbf{v}, \mathbf{h}) = h^T \mathbf{W} \mathbf{v} - c^T \mathbf{v} - d^T \mathbf{h}$$

c, d represents bias terms in the above expression. The system's 'state' probability is determined by transferring the E exponentially below and taking normalization.

$$P(\mathbf{v}, \mathbf{h}) = \frac{1}{Z} [e^{\{-E(\mathbf{v}, \mathbf{h})\}}].$$

Partitioning feature Z is typically stubborn to calculate. Nevertheless, contingent inference is manageable in the context, v-conditioned h computation and ends with the following expression:

$$P(h_j \mid \mathbf{v}) = \frac{1}{[1 + e^{\{-b_j - \mathbf{W}_j \mathbf{v}\}}]}$$

Because of system symmetry, $P(v_i \mid \mathbf{h})$ will have a similar expression.

To differentiate healthy lung tissue from fibrous and other types, Van Tulder [33] changed RBMs into so-called convolutionary RBMs. Their dataset was composed of 128 patient's CT chest scans being affected by diffuse parenchymal lung disease (DPLD) taken from a database called interstitial lung disease (ILD). They trained the Convolutional RBMs to learn features with either fully generative, fully discriminative or combined generative and discriminatory learning objectives. Depending on the proportion of generative learning and the size of the input patch, classification accuracies from 41% to 68% were achieved. The best results were achieved by mixed generative and discriminatory learning. From these results, they concluded that unsupervised feature extractors could be facilitated by discriminative learning to acquire filters tailored for categorization objectives.

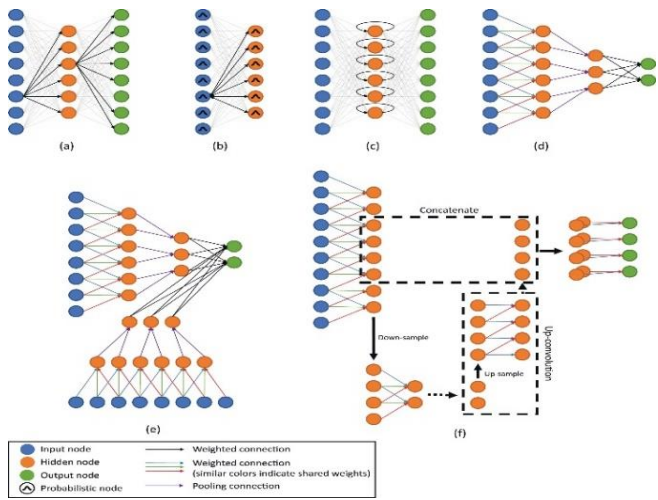

Fig 2: *Node graphs of 1D symbols usually utilized in medical imaging. (a) AE, (b) RBM (c) RNN (d) CNN (e) multi-stream CNN, (f) U-net (with a single down sampling phase). Figure from [2]*

Using Contrast-Divergence algorithms [34], the training of RBMs can be effectively done, and RBMs can be mounded into Deep Belief Networks (DBNs). Here, the output of the RBM's hidden layer becomes visible layer input of a second RBM mound on it. In 2006, Hinton et al.[35] identified DBNs, which were mostly responsible for the DL revival. Hinton et al.'s observation were that greedy, layer-by-layer type of training used for DBNs [36]. In greedy training, low and high-level features are gradually learned by lower and higher layers respectively. Additionally, DBNs can be combined with supervised RBMs to construct a semi-supervised framework in DL. Khatami et al.[37] reported an application of RBMs to group X-ray images into groups of five anatomical orientations.

Figure 2 shows the node graphs of 1D symbols usually utilized in medical imaging for some supervised and unsupervised learning Models.

## 3. DL USES IN MEDICAL IMAGE ANALYSIS

### 3.1 Classification

Classification of medical images is further divided into classification of images or exam and classification of the lesion.

#### 3.1.1 *Classification of Image or Exam*

One of the first fields where DL was applied to the study of medical images was classification. In exam classification, several images (exams) are provided to the system as input. The output of the system is a single variable classifying the image as diseased or not diseased. In this kind of environment, every diagnostic test is a sample, and the dataset sizes are small (hundreds/thousands) compared with the general computer vision (millions). Therefore, transfer learning has been successfully used for medical applications of DL. In Transfer Learning, we train our system using natural (general) images available in large datasets, to fulfil the supposed necessity for deep networks training of large data sets. Two approaches the transfer learning have been acknowledged:

(i) Extraction of features by using a pre-trained network
(ii) Refinement of the pre-trained network using medical images.

The earlier approach has the additional advantage that we do not need any training for the DL network at all so that the derived features can conveniently be incorporated into the current image processing pipelines. Both of these approaches are common and are extensively used in the classification of medical data. Nonetheless, few writers conduct a thorough study into which technique yields the best outcome. The two studies that do,[38, 39] offer contradictory results. In a paper published by Antony et al. (2016), feature extraction was outperformed by fine-tuning. The accuracy in knee osteoarthritis was 57.6% for fine-tuning relative to 53.4% for multi-class rank evaluation [39]. In the second study, Kim et al. (2016a) revealed that the accuracy was improved from 69.1% by fine-tuning to 70.5% by feature extractor for cytopathology image classification [38]. Two studies [40, 41] achieved almost human expert performance when a pre-trained variant of *Google's Inception V3* model was fine-tuned using medical data. Such findings have not yet been obtained merely by using pre-trained networks for feature extraction.

Initially, the pretraining and unsupervised network architectures such as RBMs and SAEs were concentrated by the medical imaging community. The first articles on these analysis techniques were published in 2013 and based on neuroimaging. Two studies [42, 43] used DBNs and SAEs for Alzheimer's disease classification from scans of brain MRI. Recently, a significant shift toward CNNs has been observed. The fields of use of these approaches vary widely. Applications can be seen from retinal imaging to brain MRI and lung CT to digital pathology. Instead of using pre-trained networks, writers have frequently been building their models using scratch data recently. Experiments performed by Menegola et al. [44] for performance comparison of pre-trained networks using fine-tuning against scratch training revealed fine-tuning performed well on a small dataset of one thousand skin lesion pictures. However, these studies are too limited to drawing any general conclusions. Such experiments are too small to be used to draw any general decisions. Three studies used an architecture that leveraged specific medical data attributes: instead of using 2D convolution, 3D convolution was used by two studies for Alzheimer's patient identification [45, 46]. A CNN-like architecture was applied to a graph showing the connectivity of brain by [47], which was obtained using *Diffusion-Tensor Imaging* (DTI). Various additional layers were established by them, which laid the foundations of their network. They used their system to forecast brain's growth and presented that they surpassed current methods for measuring cerebral and motor ratings. An AI system with DL approach and optical coherence tomography was presented by [48] for classification of retinal images for early detection of retina diseases. A review on classification of the thorax, eye and skin diseases using DL approaches was presented by Serte et al. can be a useful guide for researchers in the related field [49]. To summarize, the current standard techniques for image classification are CNNs. Especially pre-trained CNNs with natural images have been shown significant performance, which challenges the accuracy of professional analysts in certain activities. In the last, researchers have demonstrated that CNNs may be modified to exploit the medical image's internal structure.

### 3.1.2 Classification of Lesion

Lesion classification typically emphasizes on classifying a particular segment of the medical image into two or more classes. Accurate classification needs both local knowledge on the type of lesions and global contextual details on the position of lesions. In generic DL architectures, this combination is not possible. To resolve this issue, several researchers have been utilized multi-stream architectures in the multi-scale model, called Multi-stream architectures. Three CNNs were employed by Shen et al., each taking a lesion patch at a different size as their input. [50]. The final feature vector was obtained by concatenating the individual output of the three CNNs.

Kawahara and Hamarneh followed a somewhat similar approach for skin lesions classification by using a multi-stream CNN. In the multi-stream CNN, different resolutions were provided to each stream at input [51]. A combination of RNNs and pre-trained CNNs was proposed to grade Nuclear Cataracts from slit-lamp lens images [52]. Their approach enables the processing of all relevant information independently of the image size. A necessity for great results in object classification challenges in medical imaging is mostly incorporating 3D details. Since natural computer vision images are mostly 2D, 3D information is not directly leveraged by networks developed in those scenarios. Researchers have already employed various methods to incorporate 3D successfully with traditional architectures. For classification of regions of interest (ROI) in two classes, nodule or non-nodule in the chest CT, multi-stream CNN were used by Setio et al [53]. Up to nine specifically directed patches extracted from the individual have been used in different layers and combined through the fully connected layers for achieving an ultimate classification output. By contrast, the 3D aspect of MRI was utilized by Nie et al. to train a 3D CNN for lifespan assessment of patients with high-grade gliomas [54]. Most researchers favor using CNNs trained from end to end. Other structures and methods, for example SAEs [55], RBMs [33, 56], and convolutional sparse auto-encoders (CSAE) are used in certain cases [24]. The main distinction between a standard CNN and CSAE is using the unsupervised pre-training through sparse auto-encoders. Combining multiple learning (MIL) and DL is a useful method, particularly in situations where the object annotation to produce training data is costly. Xu et al. explored the usage of a MIL system for both supervised and unsupervised strategies to feature training and also hand-made features [57]. The findings showed that the efficiency of the MIL system was higher than that of the handmade features. We also assume these methods to be common over time, as it is difficult to collect high-quality labeled medical data. Compare to exam classifications, usage of pre-trained networks is less for object classification, because contextual or three-dimensional information needs to be incorporated. Many scholars have identified creative ways to apply this insightful to deep networks with decent results, and we anticipate DL in future to become much more popular for this mission.

### 3.2 Detection

### 3.2.1 Localization of Landmark or Object

Localization of physical objects (in time or space), like organs or landmarks, became a necessary pre-processing phase in segmentation activities. Localization in medical imaging often requires 3D-volume parsing. Several methods have been shown to solve parsing of 3D data with DL algorithms, which treat 3D space as 2D orthogonal plane configuration. Landmarks on the distal femur surface were identified by [58] by using regular CNNs to process three separate sets of 2D MRI slices. The landmark's 3D location was described as combining the 3 2D cuts with the highest performance of classification. By defining a 3D rectangular box after 2D parsing of the 3D CT size, de Vos found ROIs around the heart's physical parts, descending aorta and aortic arch [59].

For the same reason, pre-trained CNN frameworks and RBM were used to solve the shortage of data to learn better representations of the features [60]. These studies portray the challenge of localization as a classification activity and can exploit generalized deep learning models and learning processes treated accordingly. Many researchers attempt to change the mechanism of network learning to estimate their positions directly. Payer et al. recommended that landmark positions be restored instantly through CNNs [61]. They utilized landmark plots as ground-truth input data, where a Gaussian denoted each landmark, and the network was trained to forecast this landmark map. Ghesu et al. applied reinforcement learning for landmark identification. In multiple activities, like cardiac 2D MRI and US, and 3D neck/head CT, the researchers reported significant results [62].

A few approaches tackled the direct localization of spots and regions in the 3D image domain due to its enhanced complexity. Zheng et al. [63] decomposed 3D convolution into three 1D convolutions to minimize this complexity in CT data for the identification of carotid artery bifurcation. A dynamic, sparse DL network assisted by marginal spatial learning was proposed by Ghesu et al. [64] to address data complexity in 3D transesophageal echocardiogram detection of the aortic valve.

In temporal data, CNNs were also used to locate scan planes or key frames. For detecting around 12 uniform mid-pregnancy scan planes in fetal US, video frame data was used for CNN training [65]. In addition, they utilized saliency maps to achieve an approximate position of the object of concern, like the spine and head in the scan layout. RNNs, especially LSTM-RNNs, were being used to manipulate temporal data stored in medical videos, which is again high-dimensional data. For example, Chen et al. [66] used LSTM architectures to insert temporary sequence details in U.S. fetal standard plane detection videos. An integration of LSTM-RNN and CNN was used to detect the heart's end-systole and end-diastole frames in cine-MRI. In conclusion, localization with CNNs using 2D image classification has been the most common overall technique for identifying landmarks, regions, and organs with impressive outcomes.

Some recent articles, however, improve on this idea by changing the learning method to focus on precise localization with positive outcomes directly.

### 3.2.2 Detection of Object or Lesion

Identifying objects of concern or abnormalities in images is a vital aspect of the treatment and for professionals is one of the most labor-consuming. Usually, the tasks are to locate and identify minor lesions in the whole image space. In computer-aided detection systems, there has been a long tradition of research intended to detect lesions automatically, enhance the accuracy of detection or decrease professionals reading time. The first CNN-based object detection technology was already developed in 1995, utilizing a 4-layer CNN to locate nodules in x-ray photographs [67].

Many of the object detection systems based on DL reported are also using CNNs for pixel classification, and after that, some type of post-processing is implemented to attain candidates for objects. Utilizing multi-stream CNNs, 3D, or Contextual details can also be incorporated. To integrate CT and PET data, [68] used a multi-stream CNN. Dou et al. [69] found micro-bleeds with a 3D CNN in brain MRI. Few factors vary significantly between the identification of objects and the classification of objects. A crucial argument is that since each pixel is graded, the class balance in a training environment is usually heavily skewed against the non-object class. Most non-object samples are typically simple to differentiate against, avoiding the DL approach of concentrating on the difficult samples. A system of selective data sampling was proposed by [70] in which misclassified samples were more frequently fed back into the system to concentrate on critical areas in retinal images.

Furthermore, because each pixel's classification in a sliding window style leads to the redundant estimation of magnitude, fCNNs, as described in [71] are also critical facets of an object detection system. Consequently, problems in the practical use of DP techniques for object detection are classification are almost identical. Just a few articles discuss problems related to object detection, such as class inequality or effective pixel / voxel-wise image processing.

### 3.3 Segmentation

### 3.3.1 Organ and subparts

In medical images, organs and other subpart's segmentation permits quantitative analysis of clinical parameters related to shape and volume, such as brain or cardiac analysis It's also sometimes a significant first step in computer-assisted detection pathways. Typically, the segmentation's role is described as determining the collection of voxels that constitute either the contour of the object of interest or its interior. Segmentation is perhaps the most popular topic of articles that apply DL to medical images and therefore has usually found the broadest range of methodologies, such as the creation of novel CNN-based segmentation frameworks and more extensive use of RNNs. Of these novel CNN architectures, the most popular in medical imaging is the one published by Ronneberger et al., called U-net [72]. U-net's two fundamental design novelties are the use of an equivalent number of layers for up and downsampling and combines the previously known learned up-sampling layers with skipping connections amid opposing layers of convolution and deconvolution. Using contraction and expansion of pathsfeatures concatenation is obtained. From a training point of view, this implies that U-net will process all images or scans in a single forward pass, directly results in a segmentation map, which enables U-net to consider the full context of the image, which unlike patch-based CNNs, may be an advantage. In addition, it has been stated that a complete 3D segmentation can be accomplished if we feed several 2D annotated slices to U-net from the same volume [73]. Some researchers have already constructed U-net architecture derivatives. One example is Milletari et al., who suggested a version of U-net in 3D, called V-net, that does segmentation of 3D images using three-dimensional convolution layers, the objective function of which is dependent on Dice coefficient [74]. Apart from the lengthy skip-connections in a standard U-net architecture, the usage of shorter ResNet-like skip connections was also examined [75].

Recently RNNs became more common for segmentation purposes. One example is the RNN used by Xie et al. for the segmentation of the H&E histopathology images [76]. This RNN considers previous info from both the current patch's column and row antecedents. The RNN is applied four times in various directions to add bidirectional details from both right/bottom and left/top peers, and the final result is provided to an FCL after concatenation. This creates the result for a solo patch. Usage of a 3D LSTM-RNN with convolutional layers in 6 directions was proposed for the first time by Stollenga et al. [20]. Bi-directional LSTM-RNNs, in combination with 2D U-net, were used by Chen et al. for structure segmentation in 3D *Electron Microscopy* (EM) [19]. To achieve 3D segmentation, Poudel et al. merged a 2D U-net design with a RGU [77].

Although these unique architectures of segmentation provided convincing benefits, several scholars have achieved outstanding segmentation outcomes with patch-trained NNs. Among the first articles with DL algorithms addressing medical image segmentation utilized, this technique was reported by Ciresan et al. [78]. In a sliding window fashion, they implemented pixel-wise segmentation of membranes in EM imagery. Recent research articles now use fCNNs to reduce redundant computation in preference to sliding window-based classification. A single fCNN was trained by Moeskops et al. for segmentation of breast MRI pectoral muscle, brain MRI, and cardiac CT *angiography* (CTA) coronary arteries [79]. One problem with methods based on voxel classification is that they often result in false replies. To overcome this, researchers have sought to combine fCNNs with schematic systems such as MRFs and Conditional Random Fields (CRFs) to optimize the segmentation output [80-82]. In most cases, schematic systems are implemented on top of the probability map generated by fCNNs or CNNs and behave as regularizers for labels.

Custom architectures were developed to address the segmentation process directly. Such findings have been positive, rivaling and sometimes enhancing the results obtained with fCNNs.

### 3.3.2 Lesion segmentation

In the application of DL algorithms, lesion segmentation combines the problems related to the detection of objects and the segmentation of organs and substructures. To conduct efficient segmentation, global and local contexts are usually required, such that multi-stream systems of various sizes or patches that non-uniformly sampled are used [83] [84]. U-net and related models have also been seen in lesion segmentation to exploit both this global and local background. Wang et al. (2015)'s design, identical to the U-net, comprises of same up sampling and down sampling routes, but skip connections have not been included [85]. Brosch et al. used another U-net-like architecture for the segmentation of lesions in brain MRI [86]. However, they utilized 3D convolutions and only one skip link between the 1st convolutionary layer and the final deconvolutionary ones. Class imbalance is another difficulty shared by segmentation of lesion and object detection because in an image most of the pixels/voxels come from the un-diseased class. Brosch et al. described it as a weighted mixture of specificity and sensitivity with a greater specificity weight to make it less susceptible to unbalance data [86]. Data augmentation was applied by some researchers on positive samples for balancing the data collection [83, 87, 88].

Segmentation of lesions, therefore, involves a combination of methods used in organ segmentation and object detection. Innovations of these 2 fields would most definitely spread naturally to lesion segmentation, as the current problems are both pretty similar.

### 3.4 Registration

In medical image analysis, medical image registration is a common task in which a coordinate transition is determined between two medical images. This is mostly achieved in an iterative system in which a particular kind of parametric transition is considered as well as a fixed metric is configured. Whereas detection and segmentation of lesions are more common themes for DL, scientists have noticed that deep networks can help obtain the best possible performance for registration. In the current literature, two strategies are predominating.

i.  Usage of DL networks to predict the resemblance between two images that support the incremental optimization approach and

ii. Deep regression networks are used to predict transformation parameters directly.

The first strategy was used for the optimization of registration algorithms [89, 90]. Two types of SAEs were utilized by Cheng et al. to evaluate the local similarity of the head's CT and MRI images [90]. Both AEs take vectorized CT and MRI image patches and recreate them across four layers. Upon pretraining the networks utilizing unsupervised patch restoration, they are fine-tuned using two predictive layers built over the SAE's third layer. Those layers of prediction decide whether two patches are identical or different. Simonovsky et al. [89] utilized the same technique to measure the cost of similarity between two patches from separate modalities, but with CNNs. They also proposed utilizing this metric's derivative to explicitly refine the transition parameters that are separated from the system themselves. Miao et al. [91] and Yang et al. [92] directly forecast the registration transform parameters from input images using DL systems. To determine the posture and location of an embedded object during operation, Miao et al. utilized CNNs to execute a 3D model to 2D registration of X-ray. The training of CNNs was done using virtual samples created by manually adjusting the parameters of transformation for input training data. They reported that their methodology has a drastically higher registration performance level than conventional registration approaches based on pure intensity. Yang used the OASIS data set to address the issue of prior brain MRI registration [92]. As a basis, they utilized the registration technique of *large deformation diffeomorphic metric mapping* (LDDMM). For each pixel, this approach uses an initial momentum value as input, which is then developed over time to achieve the end transition. They visually got the same outcomes, but the execution time was significantly improved. Unlike classification and segmentation, the research world has not yet established the best approach to incorporate DL strategies into the registration processes. Not too many good articles have been published on the topic previously and have used particular methods. Therefore, it seems unsuitable to give recommendations on which method is most talented. However, in the near future, we hope to see even more efforts of deep learning to registration of medical images.

### 3.5 Other imaging techniques

### 3.5.1 image retrieval from Content

*Content-based image retrieval* (CBIR) is a strategy for the exploration of information in vast repositories and provides the potential of detecting common case history, discovering unusual diseases which, finally, optimizing medical care. The key task in designing CBIR approaches is to derive efficient feature descriptions from information at the pixel level and associate them with realistic concepts. The feature learning ability of deep NN at multiple levels has generated inquisitiveness in the CBIR community.

Most of the existing methods are using pre-trained CNNs for the extraction of feature descriptors from medical images. Anavi and Liu utilized their techniques to X-ray image repositories. Both used a CNN with 5 layers for features extraction from the layers which were fully connected [93, 94]. To achieve the detailed feature vector, Liu utilized the second last FCL and a routine CNN trained to classify X-rays in 193 categories [95]. On binarization of the feature vector and recovery of data through Hamming measures, the result was lower than state of the art attributed by the writers to limited patch values of 96 pixels. The last layer and a pre-trained network were used by Anavi et al. [94]. Their maximum performance was achieved when those features were fed into a single-vs-all SVM classifier for achieving the distance metric. According to their findings, including details about gender lead to improved performance than merely CNN features. Shah et al. proposed a method in which hashing- forests were combined with feature descriptors of CNN. Of touching patches in prostate MRI images, thousand features have been derived and from those features a huge matrix was built over all images. Hashing forests were utilized for each volume to compact this into descriptors [96]. Thus, CBIR has not seen many effective implementations of DL techniques until now, but soon it will be there in the field. Specific training of deep models just for the retrieving process may be an important avenue of research.

### 3.5.2 Creation and Improvement of Image

Several approaches for creating and improving images utilizing deep algorithms have been developed, including the normalization of images, eliminating hindering items in images, enhancing image quality, completing data, and discovering patterns. 2D or 3D CNNs have been used in image generation to transform one reference image into the other. These frameworks usually lack the PLs contained in classification frameworks. Such models are then trained using supervised learning. The variations between the produced and actual output are called the loss function. Different examples are present in literature, like bone-suppressed and regular X-ray in [97], 7T and 3T MRI for brain presented in [98], CT from MRI in [99] and PET from MRI in [100]. Li et al. also revealed that when the original details were incomplete or not obtained, one may use such created images for Alzheimer's disease in computer-aided diagnostics systems. Multiple low-resolution inputs can be converted to Super-resolution images by using multi-stream CNNs. Multi-stream networks have recovered high-resolution cardiac MRI from single or several low-resolution MRI inputs in [101]. This strategy can not only be utilized for inducing lost spatial information but even may use in many fields.

One example is deducing improved parameters of MRI diffusion from incomplete data [102]. Other image improvement applications, such as denoising and intensity normalization, just saw the limited implementation of DL architectures. Janowczyk used SAEs for the normalization of H&E-stained histopathology images [103]. CNNs were utilized in the DCE-MRI time series to achieve denoising. Image creation has seen an impressive performance in substantially different roles, including some innovative implementations of deep networks. We can only expect a further increase in the number of roles in the future [104].

### 3.5.3 A blend of image details and reports

Combining medical image details and text reports has directed to two research areas.
i.    Taking advantage of reports to improve accuracy in image classification [105],
ii.   Using images to create text reports [21, 106, 107]

Schlegl made the initiative towards exploiting reports, suggesting that significant volumes of labeled data might be challenging to obtain, so recommend to incorporate textual details as labels from reports [105]. Together with their textual descriptions, the system was trained on image sets and was instructed through testing to identify semantic class labels. It was also shown that in *Optical Coherence Tomography* (OCT) images, the accuracy of classification can be improved by using semantic evidences for a variety of pathologies. A comprehensive data collection of radiology images and reports taken from a PACS system was investigated for semantic connections by Wang et al. [21]. Shin suggested a model for producing details from X-rays in the chest [107]. A CNN was utilized to create a one-label image representation that was later used for RNN training to produce a series of MeSH keywords. Kisilev used three descriptors, density, margin, and shape, used in mammography, each with its class label to predicts BI-RADS descriptors for each category in breast lesions [106]. The region proposals and image data for prediction of precise label for every descriptor provided to the system. Using the image data available in PACS system and their respective diagnosis reports can be an ideal area in DL in the future.

## 4. ANATOMICAL APPLICATION AREAS

DL's applications in medical imaging proliferated. DL's freedom concerning the extraction of features has rendered it an enticing resource for imaging researchers, developers and students similar. Many of the current tools that were relied on ML resources are shifting their focus to DL evermore. There are several causes for this shift. The first is the difficulty of determining suitable tissue identification features for the different types of diseases. That caused by the reality that it is not easy to collect ground-truth data because of cost reasons. Unique datasets need manual descriptions that can be costly because of the time required for manually tracing the data. Secondly, because of 3D involvement in the ground truth of histology and pathology, this becomes very difficult, very time consuming and very costly. The 2nd issue with an ML-based system is the comprehensive search technique required to extract features from various types of frameworks. Each framework offers various sets of features, and the final feature matrix results in thousands of characteristics when combined them to different frameworks. Therefore, one requires thorough feature selection strategies that positively affect speed, generalization effect and understanding.

In the following subsections, we discuss numerous practical implementations of DL in diverse fields of medicine. We illustrate several main achievements and address system performance on some datasets and the datasets of public contests.

### 4.1 Cardiac

DL has been implemented in many dimensions of cardiac image analysis. MRI may be the most researched approach and the most popular operation in left ventricle (LV) segmentation; however, applications are diverse. These include segmentation, slice classification, tracking, quality assessment of image, automated calcium scoring, super-resolution, and coronary centerline tracking. Earlier attempts that used the active shape and appearance model [108] to detect the size and contour had its limits. Many articles used basic 2D CNNs, which evaluated slice by slice the 3D and sometimes the 4D information; the only exemption is Wolterink, which used 3D CNNs [71]. Within the U-net framework, Poudel implemented a recurrent relation to fragment the left ventricle in slices and recognize which details to recall from last slices while fragmenting the next slice [77]. Kong utilized LSTM design with a typical 2D CNN for conducting temporal regression to identify unique frames and a cardiac order [109]. Most articles used publicly available datasets. The goal of the 2015's *Kaggle Data Science Bowl* was to determine volumes of end-systolic and end-diastolic automatically in cardiac MRI, 192 teams participated in prize money for $200,000, and the high-ranking sides all used DL, especially fCNN or U-net segmentation structures.

Avendi et al. developed an optimized method with DL and deformable designs for the alignment and segmentation of the LV from the MRI dataset [110]. Ghesu et al. proposed a novel concept of Marginal Space DL (MSDL) [111], where LV's location in the 3D images was determined by gradually learning deep-classifiers. The results of this paper are not enough for deciding the scheme's effectiveness. This approach was, however, the first use of DL in 3D imaging. And thus, it can be regarded as a baseline in volumetric image parsing. Coronary artery disease (CAD) is one of the most prevalent heart disorder forms and the principal death reason worldwide. A useful tool for evaluation of CAD is X-ray angiography. A DL-based technique that uses CNN to detect vessel regions in angiography images is suggested in the study [112]. Despite the low resolution and complicated background of the data, the DL-based approach showed higher accuracy results, indicating its supremacy over traditional models. However, for better performance, deeper architectures are required to be attempted. To assess its clinical effectiveness, a comparative evaluation with the latest diagnostic medical techniques is needed. Karim et al. used DL-based architecture to introduce automated plaque characterization [113]. They used CNN based strategy, consisted of 4 layers of convolution followed by 3 FC layers. The size of the dataset used in experimentation consisted of just 56 images, and thus different dataset was required to show its robustness and validity. Summary of some studies utilizing DL approaches for analysis of cardiac images is given in Table 1.

### 4.2 Brain

Identifying abnormal brain areas is a difficult job because of the combination of intensity levels and abnormality variations in the images. DNNs have been widely used in many specific research areas for brain image analysis. A wide variety of findings deal with the diagnosis of

*Table 1: Summary of some studies utilizing DL approaches for analysis of cardiac images*

| SN | Authors | DL system | Modality | Application |
|----|---------|-----------|----------|-------------|
| 1 | [59] | CNN | CT | classification |
| 2 | [71] | CNN | CT | detection |
| 3 | [114] | CNN | CT | detection |
| 4 | [115] | CNN | MRI | LV segmentation |
| 5 | [116] | CNN | MRI | segmentation; |
| 6 | [117] | DBN | MRI | LV segmentation |
| 7 | [111] | CNN | US | detection and segmentation |
| 8 | [118] | CNN | US | Auto production of textual details for Doppler US images |
| 9 | [119] | CNN | US | Segmentation |

Alzheimer's disorder and brain tissue segmentation and anatomical structures. Other essential fields are lesion diagnosis and segmentation (e.g., cancers, lesions in white matter, lacuna, micro-bleeds). Aside from a few scan-level detection approaches, other techniques learn a mapping from local patches to representations and then from representations to labels.

Nevertheless, there is a chance that the local patches may have a deficiency of details essential for the particular tasks in which anatomic knowledge is of prime importance. To counter it, Ghafoorian utilized non-uniformly sampled patches to cover a broader range by slowly reducing sampling rates in patch edges [120]. Multi-scale research and a merger of representations inside an FCL are an alternate technique employed by several communities. Although brain images in all studied experiments 3D quantities, most approaches operate in 2D by analyzing 3D volumes slice by slice. DNNs have thoroughly taken on other challenges of brain image processing. Nearly all of the above-listed methods are based on MR images of the brain. In [83], parallel use is made of two DL-based architectures to improve the estimation accuracy. Every DL-based pathway is a CNN manipulating a specific size of input that are concatenated to provide precise brain lesion segmentation. When compared, the findings of the DL approach outperformed the Random Forest approach.

The dice similarity coefficient (DSC) values are small, which renders the technique problematic for clinical usage; however, it delivers a base for DL-based brain lesion segmentation systems. A CNN design for tumor segmentation was implemented in [88]. The tumors are classified into four groups, including edema, necrosis, non-enhancement, and enhancement. Two different

CNNs of various complexity were used to detect high-grade gliomas (HGG) and low-grade gliomas (LGG). The dice similarity metric was 0.88 for the complete, 0.83 for the core and 0.77 for the enhancing segmentation scheme. Since only 339 images have been used, the CNN with a wider collection of data has room for improved training and efficiency enhancement. Summary of some studies utilizing DL approaches for analysis of Brain images is given in Table 2.

*Table 2: Summary of some studies utilizing DL approaches for analysis of Brain images*

| SN | Authors | DL System | Modality | Application |
|---|---|---|---|---|
| 1 | [121] | ANN | MRI | Classification of HC /MCI/ AD |
| 2 | [83] | CNN | MRI | Tumor Segmentation |
| 3 | [122] | CNN | MRI | Lacune detection |
| 4 | [97] | CNN | MRI | Prediction of survival |
| 5 | [91] | SAE | MRI | Visual pathway segmentation; |
| 6 | [46] | CNN | fMRI | Classification of HC /MCI/ AD |
| 7 | [123] | CNN | MRI | Segmentation of striatum |
| 8 | [124] | RNN | MRI | Tissue segmentation |
| 9 | [125] | CNN | MRI | Image construction |
| 10 | [92] | CNN | MRI | Image registration |

### 4.2 Chest and Lung diseases

The most frequently employed application is the identification, characterization, and description of nodules in thoracic image study in both radiography and computed tomography. Many projects incorporate features from deep networks to present a set of features or compare CNNs with traditional ML methods utilizing handcrafted features. Many research groups have been identified as multiple diseases in chest X-ray, using a single system. The most popular radiological test is chest x-ray; several studies have used a combination of text reports with images to train the systems that use both RNNs and CNNs, RNNs for text analysis and CNNs for image analysis. Research area that we hope to study more shortly. In LUNA16, a task was to detect nodule in CT, all high-quality systems used CNN based architectures. Nevertheless, systems that use deep networks (e.g., U-net) for detecting candidates have performed very well.

Lung cancer is a malignant condition with a mortality rate of less than 20% over five years unless treated early. As tiny lung nodules are seldom deemed malignant, therefore it is a big challenge to characterize lung disease from CT imaging, challenging to biopsy and PET scans may not efficiently described it. DL-frameworks have created a new research area in this direction. In a study by Hua et al. two DL-based structures are used for lung cancer characterization [126]. Utilizing two DL approaches like, CNN and DBN, two different reports were carried out and observed that both CNN and DBN provided higher performance than other traditional approaches. The respiratory components in the lung may be classified as malignant depending upon their shape, that recorded for CNN were 73.3% and 78.7% while for DBN was 73.4% and 82.2% respectively. This study may be considered as a ground-level study on using DL for the detection of lung involve sphericity and configuration of internal components such as fat, calcification, and fluid. CT images of the chest were obtained from 1010 patients. For increasing the accuracy, several deeper layers could be applied.

CNNs together with SVM and RF classifiers were used by [50] to discriminate malignant lung nodules from benign. They experimented on 1010 CT lung images already labeled from Lung Image Database Consortium (LIDC-IDRI). Three CNNs were utilized in parallel, each having two convolutional layers, with each CNN taking patches of pictures in various sizes for feature extraction. The created feature vector was classified into benign or malignant using either SVM or RF classifier. The accuracy of their model was 86%.

The 2017 Kaggle Data Science Bowl [7] challenge included identifying cancerous lung nodules in CT lung scans. For this contest, about 2000 CT scans were issued and a logarithmic loss score of 0.399 was achieved by the winner [127]. Their implementation used a 3-D CNN to first separate local patches for the detection of nodules. To classify the cancer this output was then fed into a second level, which was consist of two FC layers. Summary of some studies utilizing DL approaches for analysis of Chest images is given in Table 3.

### 4.4 Breast

One significant predictor of cancer is the appearance of mitotic figures in histology images. Mitosis is a complex mechanism in which the nucleus of cells undergoes different changes. In histology images, there are many structures of the same shape and intensity, among which a few are mitotic cells. Hence, it is a challenging task to recognize mitotic cells. In [128] each pixel was classified as mitosis or non-mitosis by applying DL to the dataset specified in [129]. They got good accuracy on the data set contained only 50 images. For validation, they need a larger dataset.

Table 3: Summary of some studies utilizing DL approaches for analysis of Chest images

| SN | Authors | DL System | Modality | Application |
|---|---|---|---|---|
| 1 | [132] | GoogLeNet CNN | X-ray | Pathology detection |
| 2 | [133] | CNN | CT | Segmentation |
| 3 | [134] | CNN | X-ray | Frontal/lateral classification |
| 4 | [135] | CNN | CT | Segmentation |
| 5 | [136] | CNN | CT | Interstitial lung disease |
| 6 | [137] | CNN | X-ray | Bone suppression |
| 7 | [138] | CNN, CRF | CT | Interstitial lung disease |
| 8 | [139] | CNN, SDAE | CT | Segmentation |
| 9 | [140] | MIL | X-ray | Tuberculosis detection |
| 10 | [141] | CNN | CT | Segmentation |
| 11 | [142] | CNN | MRI | Lungs cancer detection |
| 12 | [143] | CNN | CT | Lungs cancer detection |
| 13 | [144] | Alex net | CT | Interstitial lung cancer |
| 14 | [33] | DBN,RBM | CT | Interstitial lung cancer |
| 15 | [145] | CNN | CT | Lung cancer segmentation |
| 16 | [146] | CNN | CT | Liver segmentation |
| 17 | [147] | CNN | CT | Lung CT registration |
| 18 | [148] | CNN | X-ray | Lung texture classification |

Huynh, B. Q et al. used a technique for classifying mammographic images of breast lesions using CNN-based transfer learning [130]. Comparison of the classification was found between analytically derived hand-designed lesion features and deep CNN features in [128]. Using transfer learning, a deep CNN was trained on general tasks to classify breast cancer. AlexNet was used to extract features from the dataset of breast cancer composed of 219 breast lesions on mammographic images. An accuracy of 0.86 was achieved by the system. [131] developed a DL model for breast cancer detection using mammograms of different densities. Two CCN were trained by 3002 merged mammogram images taken from 1501 subjects. The mean AUC obtained was 0.952 and 0.954 by DenseNet-169 and EfficientNet-B5 respectively.

Arevalo et al. performed studies from 344 cancer patients on a standard dataset of 736 craniocaudal mammographic and mediolateral oblique views [149]. They manually segmented the data into 310 malignant lesions and 426 benign ones. After necessary improvement, the data was provided to CNN for distinguishing malignant and benign lesions. They got an AUC of 82.6%. Huynh and Giger utilized CNN on breast US data consist of 2393 ROIs from 1125 patients for features learning [150]. The SVM was used for features classification into malignant, benign and cystic and obtained good results. In their two experiments, they got 88% and 85% AUC on CNN and handcrafted features respectively. Antropova and Giger examined CNN on features obtained from non-medical data of the ImageNet dataset using transfer learning [151], 4,096 features were extracted from 551 MRI images containing 357 malignant and 194 benign. SVM was used for lesions classification as benign or malignant and got 85% AUC.

Wang and Qu experimented with 482 mammographic images from 246 women aging from 32 to 70 years affected by the tumor [152]. First of all, a median filter was used to denoise the images and then using region growth, modified wavelet transformation, and morphological operations to segment the tumor. Finally, textural and morphological features were sent to SVM and extreme learning machine (ELM) for breast tumor classification and identification. The overall error rate was 96 with SVM and 84 with ELM. Summary of some studies utilizing DL approaches for analysis of breast images is given in Table 4.

Table 4: Summary of some studies utilizing DL approaches for analysis of breast images.TS & MG stands for tomosynthesis & mammography

| SN | Authors | DL System | Modality | Application |
|---|---|---|---|---|
| 1 | [153] | CNN | MG | Detection |
| 2 | [154] | CNN | MRI | Segmentation |
| 3 | [155] | CNN | MG | Differentiate healthy cysts from malignant masses |
| 4 | [156] | CNN | MG | localization and classification of masses |
| 5 | [56] | RBM | US | Classification |
| 6 | [24] | SAE | MG | Classification |
| 7 | [157] | CNN | TS | Detection |
| 8 | [158] | CNN | MG | Mass detection |
| 9 | [159] | CNN | TS | Mass detection in tomosynthesis |
| 10 | [160] | CNN | MG | Classification |

### 4.5 Gastrointestinal (GI) Diseases

GI disorders influence all the organs implicated in diet intake, nutrient absorption, and waste product elimination. The organs of digestive phenomena include the esophagus, liver, large intestine, and small intestine. A number of illnesses and disorders such as swelling, vomiting, infections, and cancer in the GI tract impair the digestion and absorption of food [161]. Ulcers cause the upper GI tract bleeding, while polyps, cancer, or diverticulitis causes colon bleeding. Illnesses such as Crohn's disease, celiac disease, malignant and benign cancers, duodenal ulcers, intestinal congestion, and irritable bowel syndrome are found in the small intestine

Image recognition and ML play a critical role in the detection and treatment of the above disorders, ensuring doctors make useful and reliable choices on immediate care. Because of developments in Computer-Aided Diagnosis (CAD) technologies, specific types of imaging assessments are utilized to identify diseases in digestive systems. Such diagnostic experiments include endoscopy, wireless capsule endoscopy (WCE), colonoscopy, enteroscopy, X-ray scans, intraoperative enteroscopy, computed tomography, deep enteroscopy of small intestines and MRI.

Jia and Meng used a DCNN in 10,000 WCE images to identify bleeding in GI infection. The F-measure reported was around 99% [162]. A study conducted by Pei et al. concentrated primarily on determining the occurrence of bowel contractions by analyzing diameter trends and the bowel duration by calculating temporal detail [163]. They used both FCN and a combination of FCN-LSTM with small and large data collections. FCN model was trained on a large dataset composed of 50 labeled raw cine-MRI sequences, while training of FCN-LSTM was done by a limited dataset comprising five unlabeled cine-MRI sequences. The ImageNet dataset was used by Wimmer et al. [164] to acquired features; the obtained feature vector was then given to CNN SoftMax to detect and classify celiac disease utilizing duodenum's endoscopy images [165].

Zhu et al. adopted a popular approach for automated extraction of endoscopic images utilizing a CNN. For GI lesion detection and classification, the extracted feature vector was given to SVM. An accuracy of 80% was recorded on 180 lesion detection images [166]. Similarly, Georgakopoulos et al. used a hybrid approach [167]. Using CNN, fast feature extraction was used and the respective feature vector was fed to an SVM to detect inflammatory GI illness from WCE videos. The KID training set composed of labeled data with 337 inflammatory and 599 non-inflammatory images was used for the experiments. The dataset used for testing was composed of 27 normal and 27 abnormal images. They got 90% accuracy.

The research involved polyp identification in colonoscopy videos utilizing three-way representation of the image [168]. A range of CNN models have been trained in multiple scales on isolated features like color, shape, texture and time details, which improved the precise position of a polyp, and the results were concatenated for the final results. Authors reported that their polyp data set was the largest illustrated dataset compared to standard methods that reduced polyp detection duration. Another pixel and CNN-based study were provided for polyp tumor staging prediction with colonic mucosa as a goal attribute [169]. Three different experiments were conducted by Ribeiro et al. using various CNN architectures [170]. Using data augmentation, the size of the data set was increased by making different image variations. Summary of some studies utilizing DL approaches for GI Diseases is given in Table 5.

Table 5: Summary of some studies utilizing DL approaches for Gastrointestinal (GI) Diseases

| SN | Authors | DL System | Modality | Application |
|----|---------|-----------|----------|-------------|
| 1 | [171] | CNN | CT | Liver Segmentation |
| 2 | [172] | CNN | Endoscopy | Lesion classification |
| 3 | [173] | CNN | CT | Liver tumor Segmentation |
| 4 | [174] | CNN | Colonoscopy | Polyp detection |
| 5 | [95] | RBM | CT | Colitis detection |
| 6 | [175] | CNN | Endoscopy | Classification |
| 7 | [176] | CNN | MRI | Prostate detection |
| 8 | [177] | SAE | MRI | Lesion classification |
| 9 | [178] | CNN | Endoscopy | Cancer detection |
| 10 | [179] | CNN | CT | Pancreas segmentation |

## 4.6 Histopathology and Microscopy

Histological research includes the analysis of cells, cell groups, and tissues. If numerous changes arise at the tissue and tissue level, microscopic variations, features, and characteristics may be observed by microscopic image technologies and stains [180, 181]. Different steps are required for this method, like staining, sectioning, fixing and microscopic optical imaging. It may be utilized for managing various skin diseases as well as other diseases. Microscopic imaging is the primary approach for parasite identification in a contaminated blood sample with smears, a primary cause for Malaria. Ziehl-Neelsen or fluorescent auramine-rhodamine stain and smear microscopy are the standard techniques used for tuberculosis detection.

Sirinukunwattana et al. [182] implemented a DCNN diagnostic classifier in the Histo Phenotypes dataset for the nuclei of colon cancer cells using stained histological images. Because of minimal histological data, using a source CNN the features were extracted using natural images from ImageNet [183] and facial images [184] and for classification, the purpose was fed to an object CNN model. Bayramoglu and Heikkila [185] performed two experiments using a CNN model to detect interstitial lung disease and thoraco-abdominal lymph node using the transfer learning method. The CNN classifier used epithelial layers and signet ring cells in tissues to be tested for assessing gastric cancer. They also reported the mitotic breast cancer figures [186].

In Quinn et al.[187], to forecast hookworm, tuberculosis, and malaria from stool samples, sputum, blood, the researchers used shape features such as morphology and moment. Using a DCNN model for classification, automatic microscopic image analysis was conducted and recorded 100% AUC for malaria and 99% AUC for hookworm and tuberculosis. A DCNN was also used in Peixinho et al. [188] for intestinal parasites and malaria treatment in Quinn et al. al.[187]. In Xie et al.[189], DL with fCNN was used for the automated counting of cells. DCNN was utilized in [190] to detect leukemia in metaphase. Four systems were developed by Dong and Bryan [191] using CNN and SVM to detect the AUC for malaria-infected cells, classified as infected and non-infected. LeNet-5, GoogLeNet, and AlexNet, three CNN architectures were used to extract and classify automated features and recorded 96.18, 98.13 and 95.79 percent respectively. Summary of some studies utilizing DL approaches for Histopathology and Microscopy is given in Table 6.

Table 6: Summary of some studies utilizing DL approaches for Histopathology and Microscopy H&E stands for Hematoxylin and Eosin staining,

| SN | Authors | DL System | Modality | Application |
|---|---|---|---|---|
| 1 | [192] | CNN | H&E | Cell segmentation |
| 2 | [193] | CNN | fluorescent, | Nucleus classification |
| 3 | [194] | DRN | H&E | Mitosis detection |
| 4 | [195] | CNN | Dermatoscopy | Skin cancer classification |
| 5 | [139] | CNN | H&E | Segmentation of colon glands |
| 6 | [103] | SAE | H&E | Stain normalization |
| 7 | [187] | CNN | Light microscopy | Malaria, tuberculosis and parasites detection |
| 8 | [196] | CNN | Dermatoscopy | Melanoma detection |
| 9 | [19] | LSTM-RNNs | Electron microscopy, | Neuronal membrane and fungus segmentation |
| 10 | [197] | CNN | Photographs | Skin tumor classification |

Collage of certain medical imaging techniques where DL has attained revolutionary performance is given in Figure 3.

## 5.  DISCUSSION

Recent developments in DL demonstrate that computers can retrieve more image information more efficiently and precisely than before. The additional improvement and optimization of DL methods for the specifications of data and images related to medical diagnosis is still a significant and related research challenge.

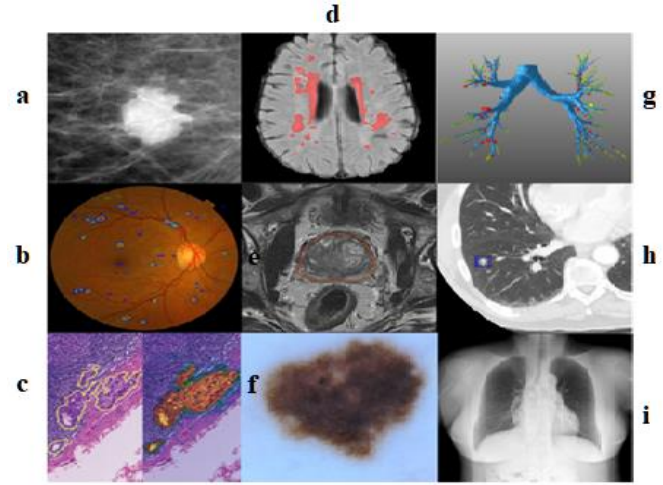

Fig 3: Collage of certain medical imaging techniques where DL has attained revolutionary performance (a) mammographic mass classification [155], (b) diabetic retinopathy classification [70], (c) breast cancer metastases detection in lymph nodes, (d) brain lesions segmentation [120], (e) segmentation of prostate, (f) human expert performance in skin lesion classification [40, 41], (g) leak detection in airway tree segmentation [122], (h) nodule classification, (i) bone suppression in x-rays [137]

### 5.1 Evaluation and robustness

The issue of small datasets is usually solved through data augmentation techniques. Data augmentation is useful, but it needs to be done appropriately. A network cannot be trained on a collection of images of a particular case and again, the testing of this professional network is done with another group of images relating to a similar issue. Likewise, treating any image slice as an individual unit may be enticing when working with 3D images. That will be incorrect because slices in a similar case are associated, and therefore, all of them should either be in the training or testing data but cannot be in both. The result would be significantly overestimated and not generalizable if not performed correctly.

By using DL for features extracting task, the enormous number of extracted features creates a problem. We would expect that the number of features won't surpass the number of data points through the usage of data augmentation, such that the selection of features or dimension reduction can be made meaningfully before any kind of classification using a particular classification model like an SVM or shallow NN. It is also necessary to remember that performance has to be measured "by case," if "case" is a subject, a lesion or something related to the therapeutic process under work. Regardless of which method we used for slicing our data, if we have a hundred subjects, the assessment must be performed according to 100 subjects only. Additionally, it is standard practice to use P-values for feature selection, but P-values are extremely variable themselves [198, 199]. Therefore, by using DL methods as extractors of features, utmost caution must be taken.

Robustness and repeatability are issues about any approach to ML [200], and so with DL. In medical image analysis, the researchers tend to reuse the same data for various tasks because the collection of medical image data sets is too challenging and are usually of small size. Correction for several comparisons is also essential to statistical performance evaluation [201, 202]. The research findings of only six out of 53 articles were confirmed by a 2012 review of 53 innovative items in fundamental cancer research [203]. In addition, a study analyzing radionics utilizing texture features showed that after adjusting P-values for several comparisons and using an appropriate cutoff (if relevant) in Kaplan–Meier investigation, all the results of 9 published studies did not achieve statistical significance [204]. Results of DL-based approaches can be much less likely to stand up under inspection if the analysis is not done correctly.

### 5.2 Datasets

Probably the most significant problem about data sets of medical imaging is to get data from a relatively large amount of correctly annotated cases. The main issue is not getting the images themselves but does getting annotations and terms of reference. The term of concern for segmentation, for example, would be the manual description of one or more expert radiologists. The primary criterion of cancer classification tasks would be the clinical reality defined by surgery or biopsy, which must be derived from diagnosis records. The term of reference must be of excellent quality, mainly if utilized for training purposes but also for evaluating the performance. It is complicated and time-consuming to acquire useful quality image data, annotations, and reference standards. Privacy laws for patients make data collection more complicated. Important information must be retrieved through radiology, pathology and other text records that seem to be laborious and possibly vulnerable to error when performed manually but not trivial when done automatically.

Another problem with medical image datasets is that, contrasting to a voltmeter or ruler, imaging systems are not measuring devices. Usually the images are not quantifiable and intended solely for human interpretation, not for computer. It is essential to examine the reliability of "conventional," and DL-based approaches in comparison to image developer or image analysis approaches. Efforts have been made to investigate the robustness of "conventional" techniques vs producer of US breast cancer treatment[205, 206], digital mammography based risk estimation of future breast cancer[207], and the features of the lung nodule [208]. However, one of the paybacks of DL-based techniques is that they can be less prone to picture differences than "conventional" approaches, leading to various manufacturers ' imaging equipment.

Another issue related to medical imaging datasets is the class imbalance, not just with DL-based approaches but also with "conventional" approaches as well. Data augmentation is one solution to mitigate the class imbalance problem in the training of DL systems. In this approach, during classifier training, we apply data augmentation only to the underrepresented class.

A small list of repositories and data sets for medical imaging is given in Table 7.

### 5.3 Data Augmentation

Based on the generative model or on current samples in a dataset, data augmentation produces new data samples. These newly generated data samples are then combined with the original data samples to maximize the variance of the data collection. In DL-based applications, data augmentation is a common procedure for raising the size of training sets these days and has shown to be highly successful for minimizing the risk of over-fitting and removing imbalances in multi-class datasets that are crucial to the achievement of generalizable models and test outcomes.

Popular data augmentation techniques commonly used in applications for medical image analysis include rotation, translation, cropping, scaling, and flipping of images [72, 209, 210]. To produce more samples for the classification of interstitial lung diseases, instead of augmenting the entire CT scans, Gao et al. arbitrarily applied jittering and cropping to patches of the original scans [140]. An image mixing technique was introduced by [211] which can impeccably insert a patch of the lesion in mammography or CT scan. In addition, these patches might be applied in many variations to the structure and features of the lesion. Even for small data sets, enhanced classification results were reported. Zhang et al. attempted for the solution of imbalanced data problem in classifying the medical images [212]. A novel augmentation technique called unified learning was proposed. The seed-labeled dataset was used for training of a specific DCNN for simultaneously obtaining image features and a similarity matrix, that could be utilized to search more analogous images to both classes of upper endoscopy and colonoscopy images.

Another kind of data augmentation comprises synthesizing images that used an object model and image formation principles of physics. The level of complexity for the image creation estimations and models can differ based on the final goal of the DL system [213]. By using a known entity's Radon transform, Yang et al. built an artificial CT dataset and modeled various exposure scenarios by applying noise to the data to train CNN for the estimation of high-dose predictions from low-dose predictions [214]. Cui et al. modeled dynamic PET data for training a stacked sparse AE based replication system for dynamic PET imaging [215]. In order to produce data for training a DL system to reduce noise in recreated CT images, Chen et al. produced noisy representations depending on the patient's CT images [216]. For registration of 2D and 3D images, only synthetic data was used by Miao et al. [217]. For DL-based localization of anatomy region using CT scans, various data augmentation techniques were compared by [218]. With an appropriate set of data augmentation, the accuracy on an independent test set was increased from 88\% to 97\% by using the same training data. They showed that different steps for data augmentation such as flips, translation, and zoom had an incremental impact on classifier efficiency. Eaton et al. used the data augmentation techniques 'mixup' proposed by[219] with its variant called 'mixmatch' for the segmentation task of medical images [220] The decision of mixing patches, made by 'mixmatch' depends on class prevalence. The performance of the segmentation task was increased with both mixmatch and mixup on BraTS dataset. For various image augmentation techniques please refer to [221]. Zhao et al. suggested a system for data synthesis to produce segmentations masks and pairs of brain MRI scans from just one labelled MRI scan. The researchers developed a hybrid spatial-intensity transition model for this task [222]. The labelled scan is distorted by the network such that the spatial layout of a given unlabeled scan is adopted. The brightness transformation network adjusts the brightness at each pixel after the layout is fixed up, so that the labeled scan looks like a given unlabeled scan. The two transformation networks together enable new labeled scans to be created from a reference-labeled scan and a number of unlabeled scans. The proposed procedure was tested in a 1-shot medical image segmentation environment, where just single labelled scan was present for training.

## 5.4 Features Interpretation

Thousands of features are extracted if a deep NN is utilized for feature extraction. These features, unlike designed handmade features, may not specifically refer to what radiologists may quickly understand. Designed handcrafted features sometimes represent something related directly to the characteristic radiologists utilize, such as the size or shape of the lesion. With features derived from deep NNs, this interpretability has almost entirely vanished. DL adoption in medical imaging can benefit from DL's achievements in other fields like auto-driving vehicles and robots. However, the usage of DL in medical imaging applications can have regulatory consequences because it would be more problematic to determine the cause precisely, when the results are incorrect, than for "conventional" applications. Unlike traditional approaches, DL systems were opaquer and provided little or no insight into their internal workings. Nevertheless, increasing attempts have been made to make DL approaches more transparent [223].

*Table 7: A small list of repositories and data sets for medical imaging.*

| SN | Name | Link | Description |
|---|---|---|---|
| 1 | TCIA | http://www.cancerimagingarchive.net | The database of cancer imaging holds a broad collection of cancer images, available for free download. There are 14355 patient's images from 77 sets. |
| 2 | ADNI | http://adni.loni.usc.edu/ | It includes image data of approximately 2000 Alzheimer's disease participants |
| 3 | OpenNeuro | https://openneuro.org https://registry.opendata.aws/openneuro | An online forum for sharing data related to neuroimaging. Comprises brain scans from 14,718 participants of different acquisition procedures and imaging modalities. |
| 4 | ABIDE | http://fcon_1000.projects.nitrc.org/indi/abide | Includes 1114 datasets with 593 controls of 521 subjects suffering from autism spectrum disorder. |
| 5 | UK Biobank | http://www.ukbiobank.ac.uk/ | Contains 15,000 subject's MRI images. Its goal is to hit 100,000. |

*Table 8: Summary of some contests for medical imaging.*

| SN | Name | Link | Description/Goal |
|----|------|------|------------------|
| 1 | Kaggle's 2018 Data Science Bowl | https://www.kaggle.com/c/data-science-bowl-2018 | Speed Cures. Spot Nuclei. |
| 2 | ISLES 2018 | http://www.isles-challenge.org/ | Using acute CT perfusion data to segment the stroke lesions |
| 3 | ISIC 2018 | https://challenge2018.isic-archive.com/ | Analysis of skin lesions for the diagnosis of melanoma. |
| 4 | BraTS 2018 | http://www.med.upenn.edu/sbia/brats2018.html | Using multimodal MRI scans to segment brain tumors |
| 5 | Grand-Challenges | https://grand-challenge.org/ | A huge number of contests in biomedical imaging. |
| 6 | CAMELYON17 | https://camelyon17.grand-challenge.org/Home | Designing algorithms for the automatic detection and recognition of metastases of breast cancer in whole-slide images of lymph node parts. |
| 7 | Kaggle's 2017 Data Science Bowl | https://www.kaggle.com/c/data-science-bowl-2017 | Lung Cancer and Machine Intelligence |
| 8 | RSNA Contest for Pneumonia Identification | https://www.kaggle.com/c/rsna-pneumonia-detection-challenge | Identify opacities on chest x rays automatically |
| 9 | Kaggle's 2016 Data Science Bowl | https://www.kaggle.com/c/second-annual-data-science-bowl | Reshaping the way we handle cardiac disease |
| 10 | MURA | https://stanfordmlgroup.github.io/competitions/mura/ | Determine Normality and Abnormality of bone from X-ray images |
| 11 | HVSMR 2016 | http://segchd.csail.mit.edu/ | Used 3D cardio-vascular MRI for blood pool and myocardium segmentation |

### 5.5 Competitive challenges

A variety of competitive challenges have appeared in the research area of medical images. During the recent past, the use of DL-based approaches has significantly increased, and DL approaches have been highest in performance during competitions for medical imaging. Litjens et al. noted in a study that it does not seem like the specific DL design is the most significant determinant in achieving a practical solution [2]. For instance, several participants utilized the same strategies and networks during the Kaggle Diabetic Retinopathy contest but achieved vastly differing outcomes. Techniques of data augmentation and preprocessing methods tend to provide a significant difference to high efficiency and robustness. How we can use the outcomes of these challenges to serve better the research of medical image analysis is still an open question. Summary of some contests for medical imaging is given in *Table 8*.

### 5.6 Experiences learned

Taking a look at the past of medical imaging, it seems like the popularity of specific approaches fluctuated over time. The achievements already accomplished with DL approaches are prominent and well-established. We assume that DL's implementation areas would develop over time as other approaches and are likely to be complemented by new methods. An essential lesson learned is that both the quality image data and annotations are fundamental, so analysis must be performed correctly. Another important lesson learned is that clinical and statistical significances are different. We know in research it is really important to determine statistical significance, but we must keep eyes on clinical significance too, and just because there's a different, newer and more complex CNN, it doesn't mean it'll always help radiologists more. Experience in the clinical task can be more beneficial than inserting further layers to a CNN, and the use of professional expertise to improve tactics, for example, innovative data augmentation or data processing approaches, are sometimes crucial for a particular clinical task.

There are also several barriers to "conventional" as well as DL-based medical image analysis approaches, including statistical and computational aspects. We need to study and strengthen the harmonization of image data, establish criteria for reporting and experiments, and provide greater access to annotated image data such as freely accessible datasets.

*Table 9: Freely accessible Codes for ML in Medical Imaging.*

| SN | Name & Reference | Implementation | Description |
|---|---|---|---|
| 1 | DLTK [224] | https://github.com/DLTK/DLTK | DL reference implementations on medical images |
| 2 | U-Net: [72] | https://lmb.informatik.uni-freiburg.de/people/ronneber/u-net | CNNN for Image Segmentation in Biomedical |
| 3 | SegNet: [225] | https://mi.eng.cam.ac.uk/projects/segnet | Semantic Pixel-Wise Labelling using a Deep Convolutional Encoder-Decoder structure |
| 4 | GANCS: [226] | https://github.com/gongenhao/GANCS | Deep Generative Adversarial Network (DGAN) for Compressed Sensing MRI |
| 5 | Graph Convolutional Networks [227] | https://github.com/parisots/population-gcn | Mixing imaging and non-imaging details for analysis of brain using Graph CNN |
| 6 | NiftyNet. [228] | http://niftynet.io | An open-source CNNs forum for analysis of medical images |
| 7 | DeepMedic [83] | https://github.com/Kamnitsask/deepmedic | Efficient Multi-Scale 3D CNN for Segmentation of 3D Medical Scans |
| 8 | V-net [74] | https://github.com/faustomilletari/VNet | 3D image segmentation |
| 9 | Brain lesion synthesis using GANs [229] | https://github.com/khcs/brain-synthesis-lesion-segmentation | Using DGANs for data augmentation in Medical images |
| 10 | Deep MRI Reconstruction [230] | https://github.com/js3611/Deep-MRI-Reconstruction | Deep Cascade of CNN and Convolutional RNN for MR Image Reconstruction |

## 6. CONCLUSION

DL is a rapidly growing area of research with a strong potential for future imaging and therapy applications. DL has now saturated nearly every part of medical imaging. DL's performance is the same or exceeded that of humans for certain non-medical activities like playing video games [231], and DL has also got achievements in a variety of applications related to medical imaging. Unlike conventional methods, DL-based approaches can replace human experts shortly. Nevertheless, many medical imaging problems are away from being solved, and the ideal DL approach and model has not yet been developed for each particular task and area of operation [232]. In addition, the combination of medical imaging approaches and additional data, like patient's age, demographics and history of diagnoses, still remains an ongoing research field that may help improve clinical decision-making results.

Three factors that can boost the DL revolution include ample data availability, computing power, and DL algorithm innovation. As mentioned above, there is a lot of recent research work aimed at addressing the issue of the small size of medical imaging datasets and some of the unique and specially developed for medical imaging, DL-based algorithms and architectures have shown great promise. A considerable number of research articles on DL-based medical imaging have been published, most over the past few years, and this trend are likely to continue ( for example [233-243]), all issued in the last two years. As practitioners and researchers of DL will get more expertise, it will become easier to identify problems according to which solution method is more reasonable: (i) best tackled utilizing end-to-end DL methods, (ii) tackled best when integrating DL with other methods, or (iii) no component of DL at all. In addition to the application of DL in medical imaging, we presume that the interest in medicine may also be leveraged to advance the overall computational approach of medical scientists and professionals, bringing *computational medicine* to the mainstream. If there are sufficiently high-impact computer science, physics, mathematics, and software systems joining the everyday routine in the health centers, adopting all such systems is expected to increase.

Moreover, the healthcare sector has not lost sight of DL's potential in medical imaging. Companies, large and small, are taking significant steps to create and launch new DL-based technologies, and major companies in medical imaging have made significant investments. We can undoubtedly say that DL's future in the field of medical imaging looks impressive.

***Consent for Publication:*** *We didn't use any personal data in our study.*

***Acknowledgment:*** *We have no funding for our review article.*